\documentclass[epj]{svjour}
\usepackage{graphics}
\begin{document}
\title{Non-diagonal flavour observables in B and Collider Physics\thanks{
  Contribution to the International Europhysics Conference on High Energy
 Physics EPS03, 17-23 July 2003, Aachen,  \mbox{Germany}, presented by W.P.}}
\author{T.~Hurth\thanks{{Heisenberg Fellow}}\inst{1} 
  \and W.~ Porod\inst{2}
}                     
\institute{ CERN Theory Division, CH-1211 Geneva 23, Switzerland,\\
SLAC, Stanford University, Stanford, CA 94309, USA \and
 Institute for Theoretical Physics, University of Zurich,\\ 
CH-8057 Zurich, Switzerland}
%
\date{CERN-TH/2003-269, SLAC-PUB-10232, ZU-TH 18/03}
%
\abstract{
Until now the focus within the direct search for supersymmetry has mainly 
been on
flavour diagonal observables. Recently lepton flavour violating
signals at future electron positron colliders have been studied. There is 
now an opportunity to analyse the relations between collider observables and 
low-energy observables in the hadronic sector.
In a first work in this direction, we study flavour violation in the 
squark decays of the second and third generations taking into account
results from $B$ physics, in particular from the rare decay 
$b \to s \gamma$.
Correlations between various squark decay modes can be used 
to get more precise information on various flavour violating parameters.
} 
\maketitle
\section{Sources of Flavour Violation}
\label{sec:1}

Within the Minimal Supersymmetric Standard Model \linebreak
 (MSSM) there are two
new sources of flavour changing neutral currents (FCNC), namely new
contributions which are induced through the quark mixing as  in the
SM and generic supersymmetric contributions through the squark
mixing. In contrast to the Standard Model (SM), the structure of the
unconstrained MSSM does not explain the suppression of FCNC processes
 that  is observed in experiments; this is the essence of the
well-known supersymmetric flavour problem.  FCNC 
processes therefore yield important (indirect)
information on the construction of supersymmetric extensions of the SM
and can contribute to the question of which mechanism ultimately
breaks supersymmetry. The experimental measurements of the rates for
these processes, or the upper limits set on them, impose in general a
reduction of the size of parameters in the soft supersymmetry-breaking
terms \cite{Hagelin:1992tc,Gabbiani:1996hi,Hurth:2003vb}.

To understand the sources of flavour violation that may be present in
supersymmetric models, in addition to those enclosed in the CKM matrix $K$, 
one has to consider the contributions to the squark mass matrices
\begin{eqnarray}
{\cal M}_f^2 &\equiv&  \\
&& \hspace*{-1cm} \left( \begin{array}{cc}
  M^2_{\,f,\,LL} +F_{f\,LL} +D_{f\,LL}           & 
                 M_{\,f,\,LR}^2 + F_{f\,LR} 
                                                     \\[1.01ex]
 \left(M_{\,f,\,LR}^{2}\right)^{\dagger} + F_{f\,RL} &
             \ \ M^2_{\,f,\,RR} + F_{f\,RR} +D_{f\,RR}                
 \end{array} \right) \nonumber 
\label{massmatrixd}
\end{eqnarray}
where $f$ stands for up- or down-type squarks.
The matrices $M_{u,LL}$ and $M_{d,LL}$  are related by
$SU(2)_L$ gauge invariance. In the super-CKM basis, where the quark mass matrices are diagonal 
and the squarks are rotated in parallel to their superpartners, the relation
reads as 
$K^\dagger M^2_{u,LL} K =  M^2_{d,LL} = M^2_Q$. In this basis the F-terms
$F_{f\,LL}$, $F_{f\,RL}$, $F_{f\,RR}$ as well as the $D$-terms $D_{f\,LL}$
and  $D_{f\,RR}$ are diagonal. All the additional flavour structure of the
squark sector is encoded in the soft SUSY breaking terms $M^2_Q$,
$M^2_{\,f,\,RR}$ (= $M^2_U$ for $f=u$ and $M^2_D$ for $f=d$) and 
$M{\,f,\,LR}^2$ (= $v_u (A^u)^*$ for $f=u$ and $v_d (A^d)^*$ for $f=d$).
Note, that the $A$-matrices are in general non-hermitian.

 These additional
flavour structures induce flavour violating couplings to the neutral
gauginos and higgsinos in the mass eigenbasis, which give rise to
additional contributions to observables in the $K$ and $B$ meson sector.
At present, new physics contributions to $s \to d$ and $b\to d$ 
transitions are strongly constrained.
In particular,  the transitions between first- and second-generation 
quarks, namely FCNC processes in  the $K$ system, are 
the most formidable tools to shape viable supersymmetric flavour models. 
As was recently emphasized again \cite{Ciuchini:2002uv}, 
most of the phenomena involving  $b \rightarrow s$ transitions 
are still largely unexplored and leave open the possibility of large 
new physics effects, in spite of the strong bound of the
famous $\bar B \rightarrow X_s \gamma$ decay which 
still gives the most stringent bounds in this sector. 
 Nevertheless, additional experimental 
information from the $\bar B \rightarrow X_s \ell^+ \ell^-$ decay at the
$B$ factories and new results on the $B_s$ -- $\bar B_s$ 
mixing  at the Tevatron might change this situation in the near future.
Within the present analysis, we take the present phenomenological 
situation into account by  setting the off-diagonal elements with an 
index 1 to 0.  
Regarding the $b \rightarrow s$ transitions, we restrict ourselves 
on the most 
powerful constraint from the  decay $\bar B \rightarrow X_s \gamma$ 
only.

Two further remarks are in order: 
within a phenomenological analysis of the constraints on the
flavour violating parameters in supersymmetric models with the most 
general soft terms in the squark mass matrices, we prefer to use  
the mass eigenstate formalism, which remains valid 
(in contrast to the mass insertion approximation) 
when the intergenerational mixing elements are not small.
Moreover, a consistent analysis of the bounds  should 
also include interference effects between the various contributions, 
namely the interplay between the various sources of flavour violation 
and the interference effects  of SM, gluino, chargino, neutralino 
and charged Higgs boson contributions.  
In \cite{Besmer:2001cj} such an  analysis was performed for the example 
of the rare decay  $\bar B \rightarrow X_s \gamma$; new bounds on simple
combinations of elements of the soft part of the squark mass matrices are
found to be, in general, one order of magnitude weaker that the bound 
on the single off-diagonal elements $m_{LR,23}$ which was derived in previous 
work \cite{Gabbiani:1996hi,Hagelin:1992tc}, where
any kind of interference effects were neglected. 

\section{Squark Decays}
\label{sec:2}

Squarks can decay into quarks of all generations if the
most general form the squark mass matrix is considered. The most
important decays modes for the example under study are:
\begin{eqnarray}
\tilde u_i &\to& u_j \tilde \chi^0_k \, , \,  d_j \tilde \chi^+_l \\
\tilde d_i &\to& d_j \tilde \chi^0_k \, , \,  u_j \tilde \chi^-_l 
\end{eqnarray}
with $i=1,..,6$, $j=1,2,3$, $k=1,..,4$ and $l=1,2$.
These decays are controlled by the same mixing matrices as the
contributions to $b \to s \gamma$. As this decay mode restricts
the size of some of the elements, the question arises as to which extent
flavour violating squark decays are also restricted. We will show below
that flavour violating decay modes are hardly constrained by current 
data. 

We will take the so-called Snowmass point SPS\#1a \cite{Allanach:2002nj}
as a specific example, 
which is characterized by $m_0=100$~GeV, $m_{1/2}=250$~GeV, $A_0=-100$~GeV,
$\tan\beta=10$ and $\rm{sign}(\mu)=1$\footnote{The SPS points are
strictly speaking defined by their low-energy parameters calculated with
ISAJET 7.58. In this letter we recalculate these
parameters using  SPheno 2.0 \cite{Porod:2003um} to include recent theoretical
developments in the calculation of the RGEs and the masses.}.
At the electroweak scale
one gets the following data: $M_2=192$~GeV, $\mu=351$~GeV, $m_{H^+}=396$~GeV,
$m_{\tilde g}=594$~GeV, $m_{\tilde t_1}=400$~GeV, $m_{\tilde t_2}=590$~GeV,
$m_{\tilde q_R} \simeq 550$~GeV, and $m_{\tilde q_L} \simeq 570$~GeV.
In the following we will concentrate on the mixing between the second
and third generations. As a specific example we have added a  set of 
flavour violating parameters
given in Table~\ref{tab:par}; the resulting up-squark masses in GeV are
in ascending order: 408, 510, 529, 542, 558 and 627 and for the down-squark
masses we find: 477, 525, 527, 533, 564 and 590 GeV.
This point is a random one out of 1000 points 
fulfilling the $b\to s \gamma$ constraint.  For the calculation of
BR$(b\to s \gamma)$ we have used  the formulas given in 
ref.~\cite{Borzumati:1999qt}. Note, that for this SPS\#1a-inspired
point both, the chargino and  the gluino loops, are important 
for the calculation of
BR$(\bar B  \to X_s \gamma)$. 
Therefore, there is an interplay between the flavour
structure of the down-type squarks and that  of the up-type squarks.

In  Tables~\ref{tab:br}, \ref{tab:brsd}  and \ref{tab:brgl} we collect
the branching ratios of squarks and gluinos that are larger than 1\%.
 In addition we also have: BR($\tilde u_3 \to \tilde u_1 Z$)=2.6\%,
 BR($\tilde u_3 \to \tilde u_1 h^0$)=1.2\%,
 BR($\tilde u_6 \to \tilde g c$) = 4\%,
 BR($\tilde u_6 \to \tilde d_1 W$)=2\%,
 BR($\tilde u_6 \to \tilde u_1 h^0$)=4.9\% and
  BR($\tilde u_6 \to \tilde u_2 Z$)=1.8\%.
Clearly all considered particles have large
flavour changing decay modes\footnote{Strictly speaking  one should use the
expression `generation violating decay modes' in this context.}.

\begin{table}
\caption{Flavour violating parameters in GeV$^2$  for our example. The
         corresponding BR$(\bar B \to X_s \gamma)$ is $4 \times 10^{-4}$.}
\label{tab:par}
\begin{tabular}{|ccc|cccc|}
\hline
$M^2_{Q,23}$ & $M^2_{D,23}$ & $M^2_{U,23}$ & $v_u A^u_{23}$ 
 & $v_u A^u_{32}$  & $v_d A^d_{23}$ & $v_d A^d_{32}$ \\
\hline
 47066 & 9399 & 46465 & 23896 & -44763 & 14470 & 15701 \\
\hline
\end{tabular}
\end{table}

\begin{table*}
\caption{Branching ratios (in \%) of $u$-type squarks for the point specified in 
         Table~\ref{tab:par}}
\label{tab:br}
\begin{center}
\begin{tabular}{|c|cc|cc|cc|cc|cc|cc|}
\hline
          & $\tilde \chi^0_1 c$ &  $\tilde \chi^0_1 t$ & 
              $\tilde \chi^0_2 c$ &  $\tilde \chi^0_2 t$ & 
              $\tilde \chi^0_3 c$ &  $\tilde \chi^0_3 t$ &
              $\tilde \chi^0_4 c$ &  $\tilde \chi^0_4 t$ &
              $\tilde \chi^+_1 s$ &  $\tilde \chi^+_1 b$ & 
              $\tilde \chi^+_2 s$ &  $\tilde \chi^+_2 b$ \\ \hline 
$\tilde u_1$ & 4.7                & 18 &
               5.2                &  9.6 & 
               $6 \times  10^{-3}$  & 0    &
               0.02               & 0    &
              11.3                & 46.4   &
               $2 \times  10^{-3}$  & 4.7 \\  
$\tilde u_2$ & 19.6               & 1.1  &
               0.4                & 17.5 &
               $2 \times 10^{-2} $ & 0  &
               $6 \times 10^{-2} $ & 0 &
               0.5                & 57.5 &
               $3 \times  10^{-3} $ &  2.9 \\
$\tilde u_3$ & 7.3                & 3.7  &
               20                 & 1.4  &
              $6 \times 10^{-2} $ & 0    &
               0.6                & 0    &
               40.3               & 3.1  &
               1                  & 18.5 \\
$\tilde u_6$ & 5.7                &  0.4  &
              11.1                &  5.3  &
               $4 \times 10^{-2} $ &  5.7  &
              0.6                 & 13.2  &
             22.9                 & 13.1  &
             0.6                  &  8.0 \\ 
\hline
\end{tabular}
\end{center}
\end{table*}

\begin{table*}
\caption{Branching ratios (in \%) of $d$-type squarks for the point specified in 
         Table~\ref{tab:par}}
\label{tab:brsd}
\begin{center}
\begin{tabular}{|c|cc|cc|cc|cc|cc|cc|c|}
\hline
          & $\tilde \chi^0_1 s$ &  $\tilde \chi^0_1 b$ & 
              $\tilde \chi^0_2 s$ &  $\tilde \chi^0_2 b$ & 
              $\tilde \chi^0_3 s$ &  $\tilde \chi^0_3 b$ &
              $\tilde \chi^0_4 s$ &  $\tilde \chi^0_4 b$ &
              $\tilde \chi^-_1 b$ &  $\tilde \chi^-_1 t$ & 
              $\tilde \chi^-_2 b$ &  $\tilde \chi^-_2 t$ &
              $\tilde u_1 W^-$ \\ \hline 
$\tilde d_1$ & 1.2                & 5.7 &
               8.4                &  30.6 & 
               $2 \times 10^{-2}$  & 1.5    &
               0.2               & 0.9    &
              16.6                & 34.1   &
               0.6  & 0 & 0 \\  
$\tilde d_2$ & 17.4               & 5.8  &
               5.1                & 15.7 &
               $7 \times 10^{-2} $ & 7.4  &
               0.3 & 09.2 &
               9.7                & 19.7 &
               0.7 &  0 & 8.8 \\
$\tilde d_4$ & 14.7                & 21.7  &
               11.3                & 2.2  &
              $5 \times 10^{-2} $ & 10.6    &
               0.5                & 8.4    &
               22.1               & 3.6  &
               1.2                  & 0 & 3.4\\
$\tilde d_6$ & 1.7                &  0.5  &
              20.5                &  6.9  &
               0.1 &  0.9  &
              1.2                 & 1.3  &
             40.3                 & 10.2  &
             3.4                  & 11.1 & 1.8 \\ 
\hline
\end{tabular}
\end{center}
\end{table*}

In \cite{Giacomo}  squark and gluino decays at the LHC
have been considered in detail
for the point SPS\#1a. It was  shown there that lepton and quark
distributions can give relatively precise information on the 
masses of the involved particles by considering the edge variables
$m^{max}_{llq}$, $m^{min}_{llq}$, $m^{low}_{lq}$, and 
$m^{high}_{lq}$  \cite{Giacomo}. These variables are kinematic variables
describing the endpoints of jet and lepton distributions in cascades
of two body
decays of supersymmetric particles \cite{Allanach:2000kt}.
Beside the assumption that flavour changing decays
are strongly suppressed, it has also been assumed in that study that
squarks of the first two generations  have approximately the same mass, 
within a few per cent. In our example the masses of the squarks 
range from 408 GeV up to 627 GeV. This feature combined with the large flavour
violating decay modes 
will give rise to additional structures in the lepton and jet distributions.
In such a case a refined analysis will be necessary to decide whether this
additional structure is  caused by background, new particles or flavour
changing decay modes.
In such a scenario a  future $e^+ e^-$ linear collider (LC) running at 1 TeV
would be of great advantage, in particular if there is some overlap of the
running times between  LHC and LC. The reason is that at the LC precise
measurements of charginos, neutralinos and sleptons are possible  
\cite{Aguilar-Saavedra:2001rg}. In addition
 $\tilde u_1$ and $\tilde d_1$ are within  the reach of 
a 1 TeV LC in our example. Feeding back the LC information into the LHC
analysis will most likely allow for an optimized exploitation of the LHC
data. 

In ref.~\cite{Hisano:2003qu} several variables have been proposed 
for extracting information on stops and sbottoms in gluino decays.
One class of these variables considers final states containing
$b \tilde \chi^+_1$. In our example, three $u$-type squarks 
contribute with branching
ratios larger than 10\%, in contrast to the 
assumption that only the two stops 
contribute.
As a consequence we expect that  additional structures
will be present in the corresponding observables. Moreover, we expect
also in this case that a combination of LHC and LC will be useful in
the exploration of these structures. A more detailed analysis 
of the relations between $B$ and Collider physics will be presented 
in \cite{newpaperHP}.

\section{Conclusions}
We have seen, that large flavour changing decays of squarks
and gluinos 
are consistent with current data from the Tevatron and the $B$ factories.
These decays will lead to additional structures in the lepton and jet
distributions, which are used to determine the edge variables proposed
for the LHC. A linear collider with sufficient energy can in principle measure
the branching ratios of the lightest up- and/or down-type squarks
proving the hypothesis of large flavour violation in the squark sector.
This information can then be put back in the analysis of the LHC data.

\section*{Acknowledgments}
W.P.~is supported by the `Erwin Schr\"odinger fellowship No.~J2272' 
of the `Fonds zur F\"orderung der wissenschaft\-lichen Forschung' of 
Austria and partly by the Swiss `Nationalfonds'.
%
\begin{table}
\begin{center}
\caption{Gluino branching ratios larger than 1\%.}
\label{tab:brgl}
\begin{tabular}{|c|c||c|c|}
\hline
Final state & BR [\%] & Final state & BR [\%] \\
\hline
$\tilde u_1 c$ & 12.9 & $\tilde d_1 s$ & 7.2 \\
$\tilde u_1 t$ & 5.7  & $\tilde d_1 b$ & 19.8\\
$\tilde u_2 c$ &  0.4  & $\tilde d_2 s$ & 6.1\\
$\tilde u_2 t$ & 7.6  & $\tilde d_2 b$ & 4.7\\
$\tilde u_3 c$ &  0.6  & $\tilde d_3 d$ & 10.0\\
$\tilde u_4 u$ & 5.5  & $\tilde d_4 s$ & 3.5\\
$\tilde u_5 u$ & 3.0  & $\tilde d_4 b$ & 4.9\\
               &       & $\tilde d_5 d$ & 2.1\\
\hline
\end{tabular}\end{center}
\end{table}

\end{document}